\documentclass{article}

     \usepackage[final]{neurips_2019}

\usepackage[utf8]{inputenc} 
\usepackage[T1]{fontenc}    
\usepackage{hyperref}       
\usepackage{url}            
\usepackage{booktabs}       
\usepackage{amsfonts}       
\usepackage{nicefrac}       
\usepackage{microtype}      
\usepackage{ dsfont }
\usepackage{xcolor}
\title{Privacy preserving Neural Network Inference on Encrypted Data with GPUs}

\author{%
  Daniel Takabi\\
  Department of Computer Science\\
  Georgia State University\\
  Atlanta, GA 30308 \\
  \texttt{takabi@gsu.edu} \\
   \And
  Robert Podschwadt\\
  Department of Computer Science\\
  Georgia State University\\
  Atlanta, GA 30308 \\
  \texttt{rpodschwadt1@student.gsu.edu} \\
   \And   
   Jeff Druce \\
   Charles River Analytics Inc. \\
   Cambridge, MA 02138 \\
   jdruce@cra.com \\
   \And
   Curt Wu \\
   Charles River Analytics Inc. \\
   Cambridge, MA 02138 \\
   cwu@cra.com \\
   \And
   Kevin Procopio \\
   Charles River Analytics Inc. \\
   Cambridge, MA 02138 \\
   kProcopio@cra.com \\
}

\begin{document}

\maketitle

\begin{abstract}
Machine Learning as a Service (MLaaS) has become a growing trend in recent years and several such services are currently offered.
MLaaS is essentially a set of services that provides machine learning tools and capabilities as part of cloud computing services. In these settings, the cloud has pre-trained models that are deployed and large computing capacity whereas the clients can use these models to make predictions without having to worry about maintaining the models and the service.
However, the main concern with MLaaS is the privacy of the client's data.

Although there have been several proposed approaches in the literature to run machine learning models on encrypted data, the performance is still far from being satisfactory for practical use.
In this paper, we aim to accelerate the performance of running machine learning on encrypted data using combination of Fully Homomorphic Encryption (FHE), Convolutional Neural Networks (CNNs) and Graphics Processing Units (GPUs).
We use a number of optimization techniques, and efficient GPU-based implementation to achieve high performance. 
We evaluate a CNN whose architecture is similar to AlexNet to classify homomorphically encrypted samples from the Cars Overhead With Context (COWC) dataset. To the best of our knowledge, it is the first time such a complex network and large dataset is evaluated on encrypted data. Our approach achieved reasonable classification accuracy of 95\% for the COWC dataset. In terms of performance, our results show that we could achieve several thousands times speed up when we implement GPU-accelerated FHE operations on encrypted floating point numbers.

\end{abstract}

\section{Introduction}

Deep learning algorithms based on Neural Networks (NN) have achieved remarkable results and empowered a wide range of applications in different domains.
However, building these models require substantial domain expertise and computational resources which might not be available to many organizations. Therefore, there has been an evolving trend toward using cloud services for this purpose. With increasing growth of cloud services, several Machine Learning as a Service (MLaaS) are offered where training and deploying machine learning models are performed on cloud providers' infrastructure. The machine learning models are deployed on the cloud and users can use these models to make predictions without having to worry about maintaining the models and the service.
However, organization may be reluctant to use MLaaS as it requires access to the raw data which is often privacy sensitive and can raise security and privacy concerns.
Our work is focused on addressing these issues and the main objective is to provide secure privacy preserving machine learning for such scenarios in an efficient and scalable manner.

There have been increasing recognition of this important issue in recent years and several approaches have been proposed. Most of the proposed approaches can be categorized into three methods: Trusted Computing Base (TCB), Secure Multi-Party Computation (SMC) and Fully Homomorphic Encryption (FHE). 
TCBs such as Intel Software Guard Extensions (SGX) or AMD Secure Encrypted Virtualization (SEV) provide hardware-based primitives that enable us to execute private data and/or code in shielded execution environments, called enclaves.
SMC protocols are algorithmic solutions based on cryptography techniques that jointly evaluate a function between multiple parties without revealing any party's private data. FHE allows one to evaluate a function on encrypted data and get the encrypted result without decryption at any point during the computations. 

TCB-based approaches are suitable for the cloud computing environments and are efficient [8]. However, it has been shown that several attacks can be devised to allow unprivileged programs to extract content from memory that should only be accessible to privileged programs [9]. 
SMC-based approaches are interactive in nature and require a large amount of communications between the participating parties, hence, they have huge  communication overhead [6, 7]. FHE-based approaches are similar to SMC-based approaches in that both are based on cryptography techniques but FHE-based approaches are non-interactive [1-4]. FHE-based approaches, however, require a large amount of computation and are considered computionally expensive. The main drawback of these FHE-based approaches is the computational overhead. For instance, CryptoNets required 570 seconds to evaluate a FHE-friendly model on encrypted samples from the MNIST dataset [1]. Although there have been some improvements in this areas including our prior work [2-4], most of the work on FHE-based approaches has been limited to small datasets such as MNIST and CIFAR-10 [1, 4, 14]. In this work, we focus on accelerating the performance of running CNNs on encrypted
data and making it practical for larger datasets and more complex networks. Our main contribution is reducing the overall computational overhead of FHE-based approaches using GPU-accelerated implementation.

In our previous work, we have developed privacy preserving MLaaS and have demonstraed its feasibility. We have provided theoretical foundation to show that it is possible to find lowest degree polynomial approximation of a function within a certain error range and provide an approach to generate those approximations [3, 4]. We have also implemented CNNs over encrypted data and shown its applicability using datasets such as MNIST and CIFAR-10 [4]. Building upon on prior work, in this work, we show that privacy-preserving deep learning is not only possible on GPUs but can also be significantly accelerated and offers a way towards efficient scalable MLaaS.

\section{Background}

\subsection{Fully Homomorphic Encryption}
FHE schemes allow for computing arbitrary circuits over encrypted data [5]. FHE schemes achieve this by taking a somewhat homomorphic encryption (SwHE) scheme that homomorphically evaluate its own decryption function and some additional operations. This is necessary because during the encryption process some noise is added to the data. This noise grows as more operations are performed on the ciphertext. If the noise grows past a certain threshold, the ciphertext can no longer be decrypted correctly. FHE schemes deal with this noise growth by using bootstrapping [5] which resets the noise by homomorphically evaluating the decryption function. The main problem is that bootstrapping is computationally expensive. Another approach of dealing with the noise growth is by constructing a so called leveled fully homomorphic encryption scheme (LFHE). Here the parameters are chosen so that a certain number of computations can be performed before the noise passes the threshold. This is only feasible for applications where the number of operations or an estimate is known beforehand.

There are a number of FHE crypto schemes and libraries available. An important shortcoming for most of them is that they only support integers. CKKS, recently presented by Cheon et al. [4], allows for computation over real numbers. Its security is based on the ring learning with error problem. CKKS supports multiplication and addition of encrypted cihpertexts. 
To encrypt a plaintext $p$ it must first be encoded. The encoding is given as $m(X) \in \mathcal{R} := \mathds{Z}[X]/(X^{n+1})$ with $m(X)= \lfloor\Delta \cdot \phi^{-1}(p) \rceil \in \mathcal{R}$ where $\Delta > 1$ is the scaling factor. To encrypt the encoded plaintext, we need to sample some error polynomials and add them to the plaintext polynomial. With an initial modulus $q$ the secret key can be chosen as $sk=(1,s) \in \mathcal{R}^2_q $; $s$ being sampled from a distribution over $\mathcal{R}$. For the public key we first need an error distribution $\mathcal{X}_e$ over $\mathcal{R}$.  Using $a$ sampled uniform randomly from $\mathcal{R}^2_q$ and $e$ sampled from $\mathcal{X}_e$, the public key can be constructed as $pk = (b=-a \cdot s + e,a)\in \mathcal{R}^2_q$. To encrypt an encoded plaintext $m$, we need to sample two error polynomials $e_0, e_1$ from an error distribution over $\mathcal{R}$. The encryption is given as $ctxt=( c_0 = r \cdot b + e_0 + m, c_1 = r \cdot a + e_1 )\in \mathcal{R}^2_q$. On the encrypted values we can perform addition and multiplication. During these operations the scaling factor is updated and later used during decryption to rescale the decoded values correctly.
\subsection{Polynomial Approximation in Neural Networks}

Neural networks rely on activation functions to introduce non-linearity. One of the most common activation functions is ReLU which is given as $f(x)=max(x,0)$. Other popular functions include Tanh and Sigmoid. The problem with all these functions is that they can not be computed over encrypted data since they rely on operations such as comparison or division that are not supported by HE schemes. Xie et al. [6] introduced the idea to approximate the activation functions with low degree polynomials. We adopt Hesamifard et al. [4] approach which improves on the idea by approximating the derivative of the activation function and integrate over the approximation to get the a substitute activation function. 


\section{Experimental Results}
In this section, we present results of implementing neural networks over encrypted data. We used IBM's HElib library [12] and Cheon-Kim- Kim-Song (CKKS) [13] encryption scheme for implementation.

\subsection{Network Architecture and Dataset}
The data we use in our experiments is from the Cars Overhead With Context (COWC) dataset [11]. It consists of 32 by 32 color images which represent an overhead view containing a car or not. The training set includes 911924 instances and the test set has 76754 instances. We train a CNN inspired by AlexNet [10] but we make some adjustments to account for input size. As shown in Table \ref{architecure1}, the network consists of four convolutional layers, 4 average pooling layers and 3 fully connected layers. Zero padding layers where included where necessary. We add an activation layer after every convolutional layer and fully connected layer. The activation after the last fully connected layer is Sigmoid. All other activation layers use a polynomial of degree two  $0.000469841857369822x^2 + 0.500000000000008x$ that approximates ReLU. 

\begin{table}
\begin{center}
   \caption{Architecture of our CNN for the COW-C dataset.}
  \begin{tabular}{ | l | l | l | }
    \hline
    Layer       &  Type &  Parameters   \\ \hline
    1  & 2D Convolutinal &  96 filters, 11x11 size, 1x1 stride, padding \\ \hline
    2  & Activation &   \\ \hline
    3  & Average Pooling 2D & pool size 2x2  \\ \hline
    4  & 2D Convolutinal &  256 filters, 5x5 size, 1x1 stride, padding \\ \hline
    5  & Activation &   \\ \hline
    6  & Average Pooling 2D & pool size 2x2  \\ \hline
    7  & Zero Padding 2D & symmetric 1 pixel padding  \\ \hline
    8  & 2D Convolutinal &  384 filters, 3x3 size, 1x1 stride, padding \\ \hline
    9  & Activation &   \\ \hline
    10  & Average Pooling 2D & pool size 2x2  \\ \hline
    11  & Zero Padding 2D & symmetric 1 pixel padding  \\ \hline
    12  & 2D Convolutinal &  384 filters, 3x3 size, 1x1 stride, padding \\ \hline
    13  & Activation &   \\ \hline
    14  & Zero Padding 2D & symmetric 1 pixel padding  \\ \hline
    15  & Average Pooling 2D & pool size 2x2  \\ \hline
    16  & Fully Connected &  4096 units \\ \hline
    17  & Activation &   \\ \hline
    18  & Fully Connected &  4096 units \\ \hline
    19  & Activation &   \\ \hline
    20  & Fully Connected &  1 unit \\ \hline
    21  & Sigmoid Activation &  \\ \hline
  \end{tabular}
   \label{architecure1}
\end{center}
\end{table}

\subsection{Results: Neural Network Inference over Encrypted Data}

Once the polynomial approximations are calculated for ReLU, we replace the activation functions in the neural network with polynomial approximations and analyze the performance of the neural network.
We first train the network architecture described in Table \ref{architecure1} using GPU implementation.
We split the dataset into 80\% training and 20\% validation.
We then train the network for 12 epochs, with 4 different scenarios: the original network with only ReLU activation functions, the network with only polynomial activation functions, the network that switches from ReLU to polynomials at the 8th epoch, and the network with uses approximations for mean pooling in addition to activation functions. We achieved accuracy of 99.71\% for ReLU only, 97.94\% for polynomials only, and about 95\% for the network with approximations for activation functions and mean pooling layer. These results confirm the network performance for HE-compatible structures.

Once the network is trained, we test it on unseen encrypted test data. We run the test data on our model encrypted and unencrypted to get comparable time measurements. As expected, running over encrypted data comes with heavy computational cost. 
We ran several encrypted images through the network and measured the average inference time. For this test, we used a machine with 56GB of RAM and an Intel(R) Xeon(R) E5-2623 v3 at 3.00GHz CPU.
For an 8 image batch, it took 4333.2 seconds over encrypted data for classification which is about 9 minutes per image. We can potentially use larger batch sizes which will in turn improve the time per image but that requires larger memory size.

\subsection{GPU Results}
In order to improve performance of the proposed approach, we investigated and developed means of performing GPU-accelerated HE operations on encrypted floating point numbers.
Our approach for doing this can be summarized as: (1) convert floating point numbers to an array of integers; (2) encrypt those integers; and (3) use Boolean circuits to perform arithmetic operations on the encrypted arrays.
We measured several timings including: (1) number of Boolean circuits required for each arithmetic operation; (2) the overall timing for each operation for both CPUs and GPUs; and (3) reduction in time by adding GPUs.
We used a GPU Titan V with 80 cores for all experiments and we were able to achieve more than 10000 times speed up on add operation and more than 4500 times speed up on multiply operation.
We also performed experiments with different number of GPU cores (10, 20, 40) to examine its effect of performance; the results show that we get near linear speed up by increasing the number of GPU cores.

Our goal was to validate that GPU-enabled frameworks achieve significant speed increases over their corresponding CPU counterparts which was confirmed by our experiments. We are currently working on extending our GPU-based implementation to include the entire FHE-compatible CNN structures.


\section{Conclusion}

In this paper, we presented an FHE-compatible CNN that is able to classify the homomorphically encrypted images for a complex CNN and a large dataset.
The main goal was to show that privacy-preserving deep learning with FHE can be significantly accelerated with GPUs. 
We implemented a CNN architecture similar to AlexNet and performed experiments with the Cars Overhead With Context (COWC) dataset. To the best of our knowledge, it is the first time such a complex network and large dataset are evaluated on encrypted data. Our approach achieved accuracy of 95\% and our results show that we could achieve more tan 4500 times speed up when we implement GPU-accelerated FHE operations on encrypted data.

\section{Acknowledgement}
This material is based upon work partially supported by the Office of the Secretary of Defense (Acquisition, Technology, and Logistics) under Contract No. HQ003419P0065.
\section*{References}

\small
[1] N. Dowlin, R. Gilad-Bachrach, K. Laine, K. L. M. Naehrig, and J. Wernsing, ''Cryptonets: Applying neural networks to encrypted data with high throughput and accuracy," Technical Report MSR-TR-2016-3, 2016. 

[2] Ehsan Hesamifard, Hassan Takabi, and Mehdi Ghasemi. 2017. CryptoDL: Deep Neural Networks over Encrypted Data. CoRR abs/1711.05189 (2017).

[3] Ehsan Hesamifard, Hassan Takabi, Mehdi Ghasemi, and Rebecca N. Wright. 2018. Privacy-preserving Machine Learning as a Service. PoPETs 2018, 3 (2018), 123–142.

[4] Ehsan Hesamifard, Hassan Takabi, and Mehdi Ghasemi. 2019. Deep Neural Networks Classification over Encrypted Data. In Proceedings of the Ninth ACM Conference on Data and Application Security and Privacy (CODASPY '19). 

[5] Craig Gentry. 2009. Fully homomorphic encryption using ideal lattices. 169–178.

[6] P. Mohassel and Y. Zhang, "SecureML: A System for Scalable Privacy-Preserving Machine Learning," 2017 IEEE Symposium on Security and Privacy (SP), San Jose, CA, 2017, pp. 19-38.

[7] Chiraag Juvekar, Vinod Vaikuntanathan, and Anantha Chandrakasan. 2018. GAZELLE: A Low Latency Framework for Secure Neural Network Inference. In 27th USENIX Security Symposium (USENIX Security 18). USENIX Association, 1651–1669.

[8] Olga Ohrimenko, Felix Schuster, Cédric Fournet, Aastha Mehta, Sebastian Nowozin, Kapil Vaswani, and Manuel Costa. 2016. Oblivious multi-party machine learning on trusted processors. In Proceedings of the 25th USENIX Conference on Security Symposium (SEC'16), Thorsten Holz and Stefan Savage (Eds.). USENIX Association, Berkeley, CA, USA, 619-636.

[9] Guoxing Chen, Sanchuan Chen, Yuan Xiao, Yinqian Zhang, Zhiqiang Lin, and
Ten H. Lai. 2018. SgxPectre Attacks: Leaking Enclave Secrets via Speculative
Execution. CoRR abs/1802.09085 (2018). arXiv:1802.09085 http://arxiv.org/abs/ 1802.09085

[10] Alex Krizhevsky and Geoffrey Hinton. 2009. Learning multiple layers of features from tiny images. Technical Report. Citeseer.

[11] Mundhenk, T. N., Konjevod, G., Sakla, W. A., \& Boakye, K. (2016, October). A large contextual dataset for classification, detection and counting of cars with deep learning. In European Conference on Computer Vision (pp. 785-800). Springer, Cham.

[12] HElib, https://github.com/homenc/HElib

[13] Cheon J.H., Kim A., Kim M., Song Y. (2017) Homomorphic Encryption for Arithmetic of Approximate Numbers. In: Takagi T., Peyrin T. (eds) Advances in Cryptology – ASIACRYPT 2017. ASIACRYPT 2017. Lecture Notes in Computer Science, vol 10624. Springer.

[14] Ahmad Al Badawi, Jin Chao, Jie Lin, Chan Fook Mun, Sim Jun Jie, Benjamin Hong Meng Tan, Xiao Nan, Khin Mi Mi Aung, Vijay Ramaseshan Chandrasekhar: The AlexNet Moment for Homomorphic Encryption: HCNN, the First Homomorphic CNN on Encrypted Data with GPUs. CoRR abs/1811.00778 (2018)

\end{document}